\documentclass[fleqn,twoside]{article}
\usepackage{espcrc2}

\usepackage{graphicx}

\newcommand{\cnx}{\textsc {conex}}
\newcommand{\cors}{\textsc {corsika\ }}

\title{First Results of Fast One-dimensional Hybrid Simulation of EAS Using \cnx}

\author{T. Pierog\address[fzk]{Forschungszentrum Karlsruhe,
        Institut f{\"u}r Kernphysik, 76021 Karlsruhe, Germany}\thanks{speaker, e-mail: Tanguy.Pierog@ik.fzk.de},
        M.K. Alekseeva\address[mosc]{D.V. Skobeltsyn Institute of Nuclear
Physics, Moscow State University, 119992 Moscow, Russia},
        T. Bergmann\addressmark[fzk],
        V. Chernatkin\address[nan]{SUBATECH, Universit{\'e} de Nantes -- IN2P3/CNRS
-- {\'E}cole des Mines, Nantes, France},
        R. Engel\addressmark[fzk],
        D. Heck\addressmark[fzk],
        N.N. Kalmykov\addressmark[mosc],
        J. Moyon\addressmark[nan],
        S. Ostapchenko\address{Institut f{\"u}r Experimentelle Kernphysik, University of Karlsruhe, 76021 Karlsruhe, Germany}\addressmark[mosc]\thanks{Now at $^{a}$},
        T. Thouw\addressmark[fzk], and
        K. Werner\addressmark[nan]}
       
\begin{document}

\begin{abstract}
A hybrid simulation code is developed that is suited for fast one-dimensional
simulations of shower profiles, including  fluctuations. It combines the 
Monte Carlo simulation of high energy interactions with a fast numerical
solution of cascade equations for the resulting distributions of secondary
particles. Results obtained with this new code, called \cnx, are presented
and compared to \cors predictions.
\vspace{1pc}
\end{abstract}

\maketitle

\section{INTRODUCTION}

Ultra-high energy cosmic rays (UHECR) are investigated by measuring the
characteristics of the secondary particle cascades that they induce in
the Earth's atmosphere.  The information obtained from these extensive
air showers (EAS) is used to infer the properties of the primary
particle, relying on a proper theoretical description of the cascade
processes.

The most natural way to get detailed information on the atmospheric
particle cascading seems to be a direct Monte Carlo (MC) simulation
of the EAS development, like it is done, for example, in the \cors
program \cite{cor04}. Nevertheless, for primary particles of very
high energies, this is not a viable option because of unreasonably
large calculation time required. The situation can be improved by
applying some weighted sampling algorithms, like the so-called ``thinning''
method \cite{hil08}. Although this approach allows the reduction of
 EAS calculation
times to practically affordable values, it has limitations.
The summation of particle contributions with very large weights creates
significant artificial fluctuations for EAS characteristics of interest
\cite{kobal}. 

A possible alternative procedure is to describe EAS development numerically,
based on the solution of the corresponding cascade equations \cite{ded11,hil12,cosmic}.
Combining this with an explicit MC simulation of the most energetic
part of an EAS allows one to obtain accurate results
both for average EAS characteristics and for their 
fluctuations \cite{KaM13}.
In this article we report on the development of an extensive
air shower calculation program of such a type:~\cnx. The MC treatment of
above-threshold particle cascading is treated in the standard way
and does not differ significantly from e.g. \cors procedure
\cite{cor04}. On the other hand, the numerical description of lower
energy sub-cascades is based on the solution of hadronic cascade equations,
using an updated algorithm of Ref. \cite{cosmic} and a newly developed
procedure for solving electro-magnetic (e/m) cascade equations.

The outline of the paper is as follows. Section \ref{scheme} describes
the calculation scheme and its basic procedures. In Section \ref{results}
we present examples of EAS characteristics obtained with \cnx\ and investigate
the accuracy of the predictions comparing the hybrid approach
with pure MC or numerical procedures, and with the results of the \cors
code. Finally, Section \ref{summ} contains a summary of our
work and discusses both potential applications of the program and
the prospects for its further development.

\section{SCHEME}\label{scheme}

A hybrid air-shower calculation scheme consists
of two main stages:~an explicit MC simulation of particle cascade
at energies above some chosen threshold $E_{{\rm thr}}$ (typically a factor of 
100 smaller than
the energy of the primary particle) and a solution of nuclear-electro-magnetic
cascade equations for sub-cascades of smaller energies. Both MC and
numerical parts model the same physics,
but the numerical solution to the cascade equations
is calculated only along the direction of the shower axis.
% which is defined by the arrival direction of the primary particle.

In the hadronic cascade one follows the propagation, interactions
and decays (if relevant) of (anti-)nucleons, charged pions, charged
and neutral kaons. All other types of hadrons produced in the interactions
and decays are assumed to decay at the place. Particle interactions
in the MC part are treated within a chosen high energy hadronic interaction
model (optionally \textsc{nexus} \cite{dre00} or 
\textsc{qgsjet} \cite{qgsa02}) and decays are simulated using the
corresponding routines of the \textsc{nexus} model. The same models
are applied to pre-calculate average secondary particle spectra for later
use by the numerical scheme. Optionally, below
some energy $E_{\rm low}^{\rm had}\sim100$ GeV one employs the \textsc{gheisha}
model \cite{fes16}. Ionization energy loss of charged hadrons is
approximated by the Bethe-Bloch equation.

The MC treatment of the e/m cascade is realized by means of the \textsc{egs4}
code \cite{egs4}, supplemented by an account for the Landau-Pomeranchuk-Migdal
(LPM) effect for very energetic electrons (positrons) and photons.
The numerical part is based on the same interaction processes as the
MC one, using Bethe-Heitler cross sections for bremsstrahlung
and pair production with corrections at low energy according 
to Storm and Israel,
the Klein-Nishina formula for the Compton process, accounting
for M{\o}ller and Bhabha processes as well as for positron-electron
annihilation (see, for example \cite{gai10,egs4}). Both LPM-effect
and photo-effect are neglected in the numerical part as it is not
supposed to be used in the energy ranges where the corresponding processes
give an essential contribution. Ionization losses of electrons and
positrons are described by the Bethe-Bloch formula with corrections
due to the density effect. 

In general, an individual shower is simulated as follows. One starts
with the primary particle of given energy, direction, at a given
initial position in the atmosphere. These define the shower axis along which 
the shower will be calculated using slant depth. For a primary hadron
initiating a shower
one simulates the hadronic cascade explicitly until
all produced hadrons fall below the energy threshold $E_{\rm thr}$.
The characteristics of all sub-threshold hadrons and e/m
particles (type, energy, and slant depth position) are written to
corresponding stacks to form the ``source terms''
for the numerical solution of the
cascade equations. The above-threshold e/m particles
are transferred to the \textsc{egs4} program, where the simulation
of the e/m particle cascade is performed in a similar way, 
with all sub-threshold
e/m particles being added to the e/m source term. As the next step,
the hadronic cascade at energies below $E_{\rm thr}$ is described
numerically based on the solution of corresponding cascade equations
and the initial condition specified by the source term. 
The results are discretized energy
spectra of hadrons of different types at various depth positions.
All sub-threshold e/m particles produced at this stage are added to
the e/m source term. Finally, sub-threshold e/m cascades are described
numerically based on the solution of corresponding e/m cascade equations
\begin{figure}[t]
\includegraphics[width=1.\columnwidth]{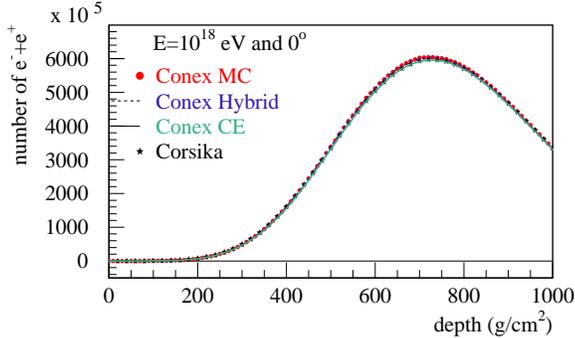}
\vspace{-1cm}
\caption{Average longitudinal profiles of charged particles of
energies above 1 MeV for proton-initiated vertical
showers of energy $E_{0}=10^{18}$ eV as calculated both in the hybrid
scheme (dashed line) and using pure MC (points) or numerical (CE)
approaches (full line), compared to \cors simulation (stars).}
\label{fig:longi}
\vspace{-0.2cm}
\end{figure}
with the initial conditions set by the corresponding source term. 
In case of the
primary particle being a photon or an electron the simulation process
consists just from e/m MC cascading for above-threshold particles
and numerical treatment of the process for sub-threshold particles.
No feed-back from the e/m to the hadronic part is considered.

\section{RESULTS}\label{results}

Currently \cnx\  can be used for the simulation of showers initiated by
any primary particle and 
energy accepted by the high energy interaction model implemented
($\gamma$, proton, 
iron and more up to $10^{22}$ eV). The basic output 
consists of the longitudinal profile along the shower axis for all produced 
particles (e/m and hadronic) except for the muons (under development). 
Additionally, Gaisser-Hillas fit parameters, longitudinal energy deposit 
profiles and energy spectra for 3 
different slant depth levels can be provided. The geometry of the model
allows the simulation of showers at any angle up to $90^{\circ}$ 
assuming US standard atmosphere, but can easily be 
adapted to any atmosphere and upward going showers. On the other hand, since 
numerical calculations are only done along the shower axis, no lateral 
distribution is available.

\begin{figure}[t]
\includegraphics[width=1.\columnwidth]{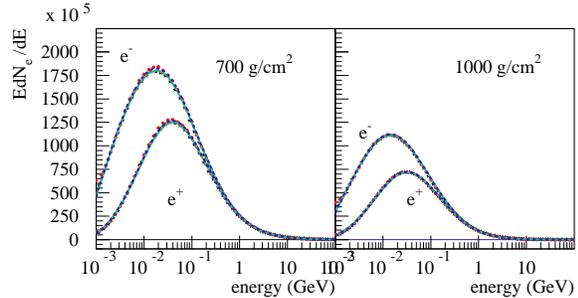}
\vspace{-1cm}
\caption{Particle energy spectra of electrons and positrons for
the depths $X=$ 700 and 1000 g/cm$^{2}$ for $10^{18}$ eV proton-initiated 
vertical showers of energy as calculated both in the hybrid
scheme (dashed line) and using pure MC (points) or numerical (CE)
approaches (full line), compared to \cors simulation (stars).}
\label{fig:energ}
\vspace{-0.2cm}
\end{figure}

In Fig.~\ref{fig:longi} the average longitudinal profile of charged particles 
for $10^{18}$ eV proton-initiated vertical showers as calculated both in 
the hybrid scheme ($E_{\rm thr}=10^{7}$ GeV) and using pure 
MC ($E_{\rm thr}=1$ GeV) or numerical ($E_{\rm thr}=10^{9}$ GeV) 
approaches with \textsc{gheisha} and \textsc{qgsjet} is
compared to \cors simulations. All methods are fully compatible with \cors 
results. Fig.~\ref{fig:energ} shows in more detail
how well
the different methods and the \cors results agree in the description
of the energy spectra of e/m particles around the maximum ($X=700$ g/cm$^{2}$) 
and at the ground level ($X=1000$ g/cm$^{2}$) for $10^{18}$ eV proton-initiated 
vertical showers.

\begin{figure}[htb]
\includegraphics[%
  width=1.\columnwidth]{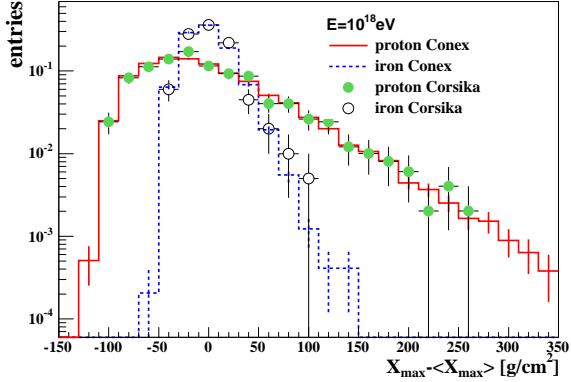}
\vspace{-1cm}
\caption{Fluctuations of the shower maximum depth $X_{\rm max}$ around
the mean shower maximum depth $\langle X_{\rm max}\rangle$
for $10^{18}$ eV proton and iron-initiated
showers simulated with \cnx\ (lines) and compared with \cors
results (points).}
\label{fig:xmax_fluc}
\vspace{-0.2cm}
\end{figure}

The most important feature for the hybrid scheme is to reproduce the 
event-by-event fluctuations of the shower development as a function of 
the primary mass. In  Fig.~\ref{fig:xmax_fluc} fluctuations of the shower 
maximum depth $X_{\rm max}$ around
the mean shower maximum depth $\langle X_{\rm max} \rangle$
for a primary energy of $10^{18}$eV are shown. Proton and iron-initiated
showers were simulated with \cnx\ and are compared with \cors
results. A perfect agreement is observed for both primary particles.

\begin{figure}[htb]
%\vspace{-.2cm}
\includegraphics[%
  width=1.\columnwidth]{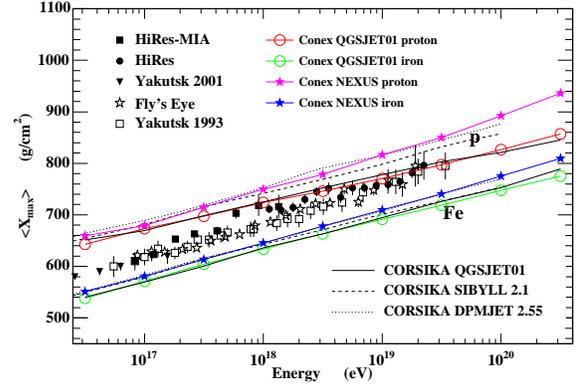}
\vspace{-1cm}
\caption{ $\langle X_{\rm max}\rangle$ of proton and iron-induced showers
 for different models together with data. \cors
shower simulations are based on 500 showers for each energy point.}
\label{fig:xmax}
%\vspace{-0.2cm}
\end{figure}

In Fig.~\ref{fig:xmax} 
the average depth of maximum, $\langle X_{\rm max}\rangle$,
of proton and iron-induced air showers is shown as
function of the energy for various models and data. The the predictions
of the \cnx\ and \cors simulation codes agree within the statistical
uncertainties.

\section{SUMMARY} \label{summ}

\cnx\ is a new, reliable tool for fast one-dimensional air shower simulations.
Realistic shower profiles, including fluctuations that
lead, for example, to showers with two maxima, can be
generated for particles of
primary mass, energy and incident angle as relevant to UHECR studies.
With a calculation time of about 2 minutes per shower (depending weakly
on the primary energy and mass), \cnx\ provides necessary information for 
applications such as fluorescence detector simulations or 
theoretical studies of shower profiles and $X_{\rm max}$
distributions. 

Work is in progress to include muon calculations and new high and 
low-energy hadronic interaction models in \cnx. Furthermore it is planned to
link \cnx\ to \cors to 
provide full 3D information and low energy particle tracking for surface 
detector analysis or shower radio emission generation.

\section*{Acknowledgments}
The work of S.O. has been supported by the German Ministry for Education
and Research (BMBF, grant 05 CU1VK1/9) and N.K.
acknowledges the support by the German Research Society 
(DFG, grant 436 RUS 17/120/03).

\end{document}